# Observation of quasi-elastic light scattering in BiFeO$_3$

Eiichi Oishi[1*], Yasuhiro Fujii[1], Akitoshi Koreeda[1], Takuya Satoh[2], and Toshimitsu Ito[3]

*[1]Department of Physical Sciences, College of Science and Engineering, Ritsumeikan University, Kusatsu, Shiga 525-8577, Japan*

*[2]Department of Physics, Tokyo Institute of Technology, Tokyo 152-8551, Japan*

*[3]National Institute of Advanced Industrial Science and Technology, Tsukuba 305-8565, Japan*

E-mail: oishi.e17@kore-lab.org

We observed quasi-elastic light scattering (QELS) in BiFeO$_3$ using Raman spectroscopy over a temperature range of 300-860 K. The QELS has two components: a narrow and broad component. The temperature dependence of the intensity and linewidth of the broad component differed below and beyond the Néel point, and the broad QELS may have a magnetic origin.

## 1. Introduction

Bismuth ferrite ($BiFeO_3$, BFO) is a multiferroic material that exhibits ferroelectricity (Curie point 1100 K[1]) and antiferromagnetism (Néel point 640 K[2]) at room temperature. The crystal structure of BFO belongs to the space group *R3c* and exhibits a spontaneous electric polarization along $[111]_{pc}$[2,3]. In addition, BFO has a G-type antiferromagnetic structure with the incommensurate spin cycloid structure derived from the antisymmetric Dzaloshinskii-Moriya interaction[4–9]. In addition to the phonons and magnons that reflect the dynamics of the crystal and magnetic structure, BFO exhibits electromagnons, which originate from the spin-lattice coupling[10,11]. These distinctive structures and dynamics have been studied via various experimental methods, such as neutron scattering and diffraction[5,12–15], X-ray diffraction[16–18], specific heat measurement[19–21], NMR[22,23], terahertz spectroscopy[24–26], pump-probe spectroscopy[27], and first-principles calculations[28–30]. In particular, there have been many reports of Raman spectroscopy[10,11,31–38] with measuring phonons and magnons. However, to the best of our knowledge, there have been no reports on quasi-elastic light scattering (QELS) of BFO.

QELS appears as a broad spectrum centered at 0 $cm^{-1}$ in the light scattering spectra of materials and has variable origins. For example, the origin of QELS can be ferroelectric polarization fluctuation[39], the fractal nature of polar nanoregions of relaxor ferroelectrics[40], phonon gas[41], and magnon gas[42]. For the QELS stem to phonon or magnon gas, its linewidth corresponds to the thermal diffusion coefficient, relaxation time of phonon-phonon scattering, spin-spin relaxation time, etc. Since the QELS of phonon and magnon gas have been reported in dielectric[43] and antiferromagnetic[42] materials, it is expected that QELS can also be observed in BFO, which is a multiferroic material. This study reports the first observation of QELS in BFO using Raman spectroscopy.

## 2. Experimental methods

We prepared a BFO single crystal with a $(111)_{pc}$-plane with a diameter of 5 mm and thickness of 1 mm using the laser-diode floating-zone method[44]. Raman spectroscopy was performed on the crystal using a 532 nm excitation source of an Nd: YAG laser and single monochromator (Jobin Yvon, HR320, grating 2400 gr/mm, resolution 0.2 $cm^{-1}$). The scattered light in the band of approximately ±5 $cm^{-1}$ was removed using notch filters[45] to prevent elastic scattering from entering the monochromator. We employed backscattering geometry and polarization configurations HH (parallel Nicol), HV (crossed Nicol), RR, and RL (Fig. 1). Here, we define the type of circularly polarized light (R or L), referring to the rotation direction of the electric field of the light at the sample surface, regardless of the propagation direction of the light. The temperature was controlled using a temperature-controlled stage for a microscope (Linkam, THMS600). The sample temperature $T$ in the irradiated region was determined from the Stokes to anti-Stokes ratio of the Raman intensity of phonon peaks. We used the following equation.

$$\frac{I_S}{I_{AS}} = \left(\frac{\omega_i - \Omega}{\omega_i + \Omega}\right)^3 \exp\left(\frac{\hbar\Omega}{k_B T}\right), \tag{1}$$

where $I_S$, $I_{AS}$, $\omega_i$, $\Omega$, $\hbar$, and $k_B$ are the Stokes and anti-Stokes Raman intensity of the phonon peaks, the angular frequencies of the incident light and phonon, Dirac's constant, and Boltzmann constant, respectively.

## 3. Results

The Raman spectra at 300 K are shown in Fig. 2, where the insets in the panels of HH and RR compare the spectrum of HH with HV and RR with RL around 0 cm$^{-1}$, respectively. The peaks with the arrows in the insets are derived from the laser. The QELS of BFO is observed as a peak at 0 cm$^{-1}$ in HH and RR, but not in HV and RL. These results suggest that the QELS is polarized. Furthermore, it appears that the QELS consists of some components because the baselines of the HH and HV (RR and RL) spectra are different. Note that the central peak in the range of -5 cm$^{-1}$ to 5 cm$^{-1}$ is suppressed by the notch filter. Because the Raman spectrum of the RR has the fewest number of phonon and magnon Raman peaks and the QELS is easily observed, we employed the RR configuration to measure the temperature dependence of QELS from 300 K to 860 K (Fig. 3(a)). Figure 3(b) shows that the QELS has narrow and broad components. We fitted the Raman spectra for each temperature in Fig. 3(a) using the Lorentz functions. The temperature dependences of the intensity and linewidth of the QELSs are shown in Fig. 4. Figure 4(b) shows that the linewidth of the narrower QELS is approximately 5 cm$^{-1}$, which is comparable to the frequency region removed by the notch filter, indicating that the temperature dependences of the intensity and linewidth of the narrower QELS are not accurate enough for us to discuss here. However, as shown in Figs. 4(c) and (d), the temperature dependences of the intensity and linewidth of the broader QELS are different below and beyond the Néel point (640 K). We discuss the origin of the broader components of the QELS.

## 4. Discussion

The broader QELS exhibits three characteristics: (1) It is polarized, (2) the intensity decreased with increasing temperature and became very small beyond the Néel point, and (3) the linewidth decreased with increasing temperature and became constant beyond the Néel point. Because the broader QELS is polarized, it is not derived from ferroelectric polarization fluctuations[39] or polar nanoregions of relaxor ferroelectrics[40], which both have depolarized QELS. The broad QELS has a magnetic origin because its intensity becomes very small beyond the Néel point. These polarized magnetic QELS could originate from the magnon gas. It has been reported that the magnon in BFO softens in its frequency with increasing temperature and disappears near the Néel point[33],[46]. Therefore, the result that the intensity of the broad QES becomes very small above the Néel point may be interpreted as the disappearance of the magnon gas. From an analogy with the phonon-gas QELS, we expect the energy-diffusion and Mountain modes to also appear in the QELS of a magnon gas. In the phonon-gas case, the linewidth of the energy-diffusion mode decreases with increasing temperature, whereas the linewidth of the Mountain mode increases with increasing temperature[41]. Therefore, the broader QELS may be attributed to the energy-diffusion mode of the magnon gas.

The Knudsen number $Kn$, which is a dimensionless quantity that determines whether the fluid can be treated as a continuum, was used to quantitatively evaluate the origin of the broader QELS. $Kn$ is defined for phonon gases as follows:

$$Kn = q \times l^{41},$$

where $q$ is the magnitude of the wave-vector transfer in the light scattering experiments and $l$ is the mean free path of phonons. In other words, $Kn$ is the ratio of the mean free path of phonons to the length scale of the fluctuations measured during light scattering. For example, for $Kn \ll 1$, that is, when $q$ and/or $l$ are sufficiently small, the phonon gas can be regarded as a continuum because there are many phonons in the observed spatial region. On the other hand, for $Kn \gg 1$, the phonon gas cannot be treated as a continuum. We assumed that the origin of the broader QELS is the energy-diffusion mode of the magnon gas. From the analogy with the phonon gas, we evaluated the $Kn$ of the broader QELS from the experimental and literature values. Because we employed a backscattering geometry, $q = 4n\pi/\lambda$, where $n$ is the refractive index of the samples and $\lambda$ is the wavelength of the light. From the phonon analogy[41], the mean free path of the magnons can be written as $l = 3\Gamma/(q^2 v)$, where $\Gamma$ is the half width at half maximum of the broader QELS and $v$ is the magnon velocity. The order of $Kn$ of the broad QELS is calculated to be $\simeq 10$, where $n$, $\lambda$, $q$, $\Gamma$ and $v$ are 3.2[47], 532 nm, $7.6 \times 10^7$ rad/m, ~$10^{12}$ Hz, and ~$10^4$ m/s[11], respectively, suggesting that magnon gas cannot be regarded as a continuum in the scattering geometry employed in this study. Therefore, the origin of the broader QELS could not be the energy-diffusion mode of the magnon gas. Although the origin of the broader QELS in BFO has not been determined in detail yet, we have shown the possibility of attributing the broader QELS to the magnetic nature of BFO for the first time.

## 4. Conclusions

We reported the QELS in BFO from 300 K to 860 K for the first time. The QELS was found to have narrow and broad components. Since the temperature dependence of the linewidth and intensity of the broader QELS differed below and beyond the Néel point at 640 K, broader QELS may have a magnetic origin. Further Brillouin and Raman measurements are essential to clarify the origins of the narrower and broader BFO QELSs.

## Acknowledgments

This work was supported by JSPS KAKENHI Grant Numbers JP19K05252, JP19H05618, JP21H01018 and JST SPRING, Grant Number JPMJSP2101.

## Figure Captions

**Fig. 1.** Polarization configurations HH, HV, RR, and RL with a backscattering geometry, where R and L represent right- and left-circular polarization. The type of circular polarization (R and L) refers to the direction of rotation of the electric field of light on the sample surface, regardless of the propagation direction of light.

**Fig. 2.** The Raman spectra of BFO in the polarization configurations of HH, HV, RR, and RL. The insets in the panels of HH and RR compare the spectrum of HH with HV and RR with RL around 0 cm$^{-1}$, respectively. The peaks with the arrows in the insets are laser derived. The QELS centered at 0 cm$^{-1}$ is observed only in HH and RR. Furthermore, it appears that the QELS consists of some components because the baselines of the HH and HV (RR and RL) spectra are different. The spectrum of RR has the fewest Raman peaks of the phonon and magnon of the four spectra.

**Fig. 3.**

(a) Temperature dependence of the Raman spectra from 300 K to 860 K. The sample temperature $T$ in the irradiated region was determined from the Stokes to anti-Stokes ratio of the Raman intensity of phonon peaks using Eq. (1). The shape of the Raman spectra changes between 570 K and 635 K near the Néel point. (b) The Raman spectrum of BFO at 545 K. The QELS has narrow and broad components.

**Fig. 4.**

(a) and (b) shows the temperature dependence of the intensity and linewidth of the narrow QELS, while (c) and (d) shows that of the broad QELS. The intensity and linewidth of the narrow QELS have almost no temperature dependence. On the other hand, the temperature dependence of the intensity and linewidth of the broad QELS changes around the Néel point.

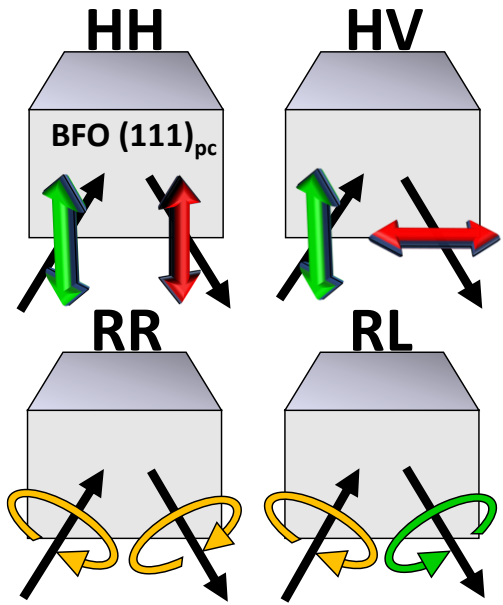

Fig.1. (Color Online)

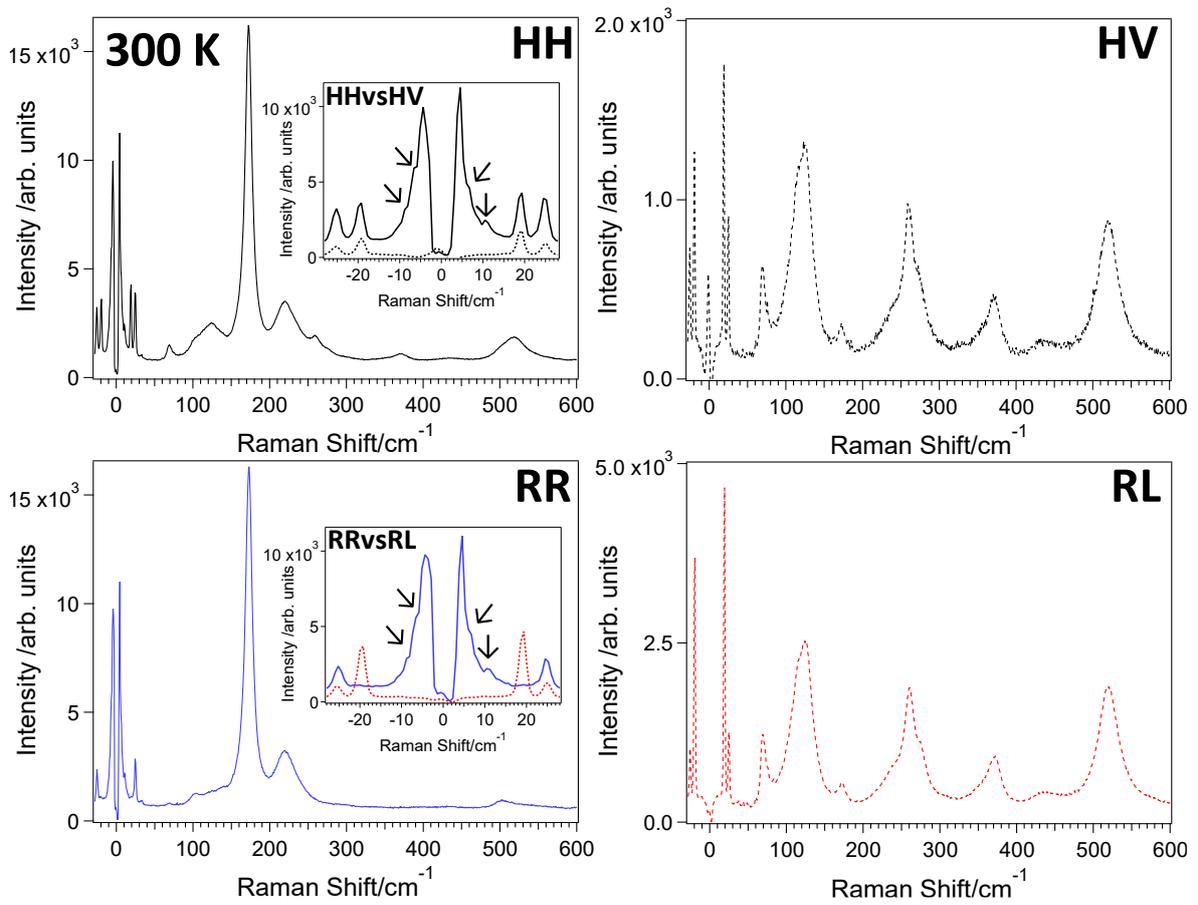

Fig. 2. (Color Online)

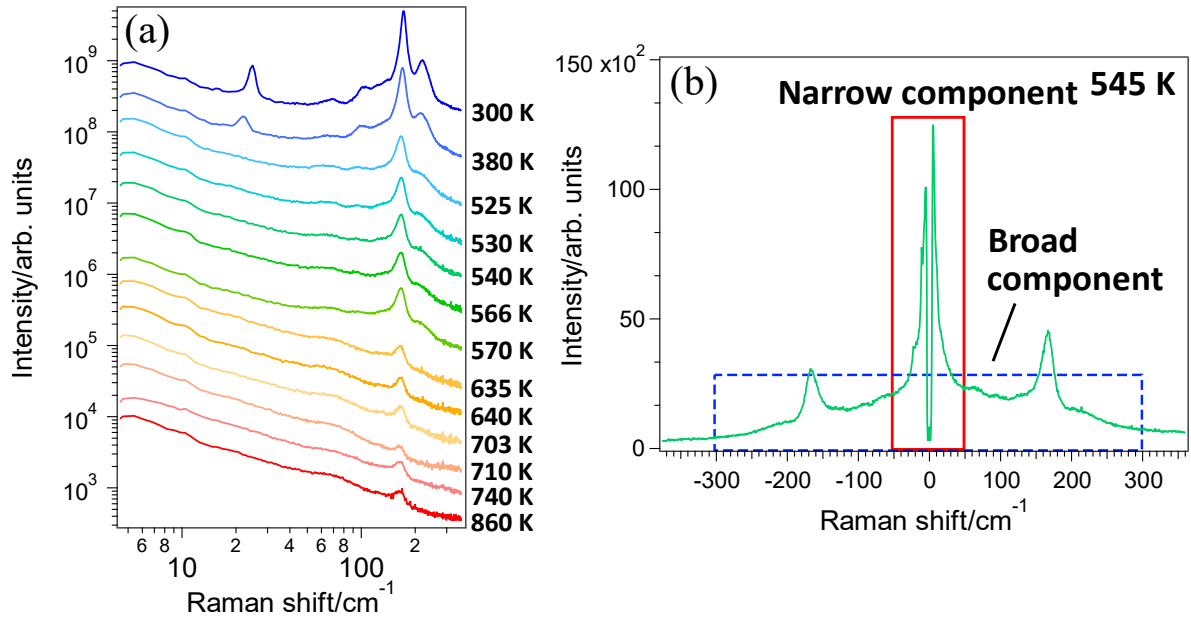

Fig. 3 (Color Online)

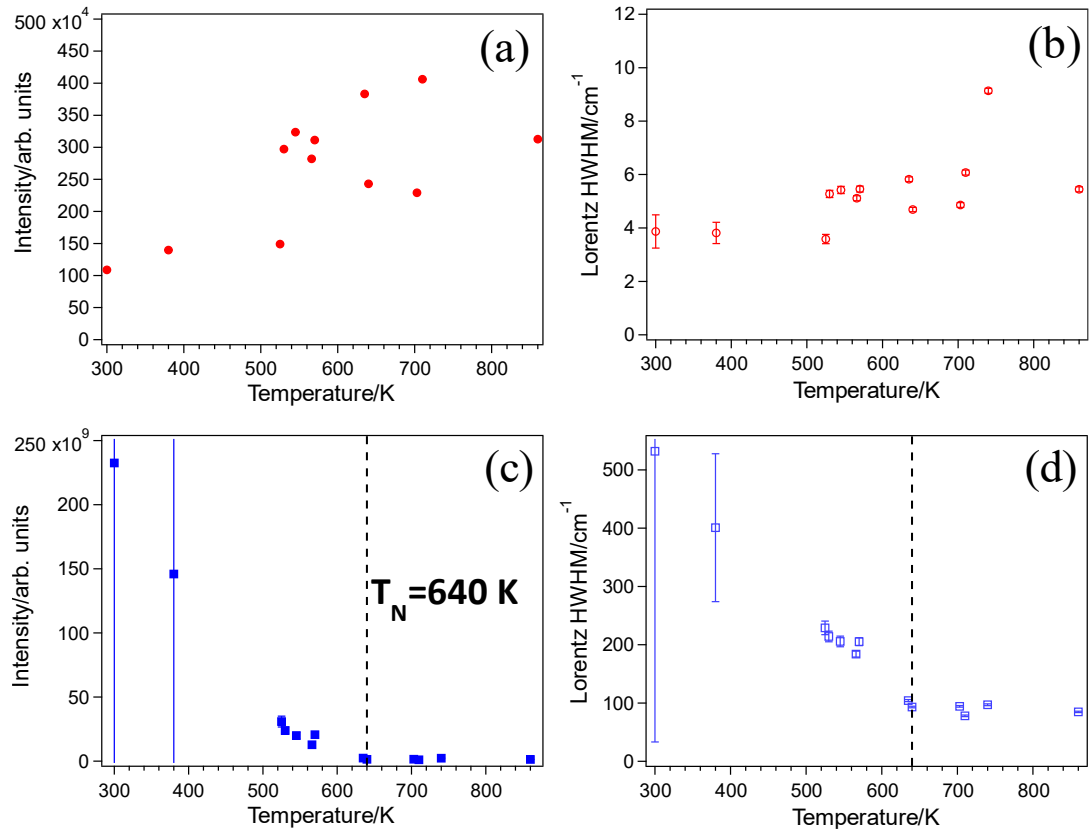

Fig. 4 (Color Online)